\documentstyle[prb,aps,floats,epsbox]{revtex}
\begin{document}
\draft 
\wideabs{
\title{Photoemission study of the skutterudite compounds \\
  Co(Sb$_{1-x}$Te$_{x}$)$_3$ and RhSb$_3$}
\author{H.~Ishii, K. Okazaki, and A. Fujimori}
\address{Department of Physics and Department of Complexity Science and 
  Engineering, University of Tokyo, Bunkyo-ku,
 Tokyo 113-0033, Japan}
\author{Y.~Nagamoto and T.~Koyanagi}
\address{Department of Symbiotic Environmental Systems Engineeing,
Graduate School of Science and Engineering, Yamaguchi University, Ube 755-8611, Japan}
\author{J.~O.~Sofo}
\address{Centro At\'{o}mico Bariloche and Instituto Balseiro, Comisi\'{o}n Nacional de Energ\'{i}a At\'{o}mica, (8400) Bariloche, RN, Argentina}
\date{\today} 
\maketitle

\newcommand{\HeI}{He \footnotesize I \normalsize}
\newcommand{\Co}{Co(Sb$_{1-x}$Te$_{x}$)$_3$ ($x=$ 0, 0.02, 0.04) }
\newcommand{\ptype}{CoSb$_3$ }
\newcommand{\med}{Co(Sb$_{0.98}$Te$_{0.02}$)$_3$ }
\newcommand{\ntype}{Co(Sb$_{0.96}$Te$_{0.04}$)$_3$ }
\newcommand{\xCo}{CoSb$_3$  ($x=$ 0) }
\newcommand{\ef}{$E_F$ }
\newcommand{\Rh}{RhSb$_3$ }

\begin{abstract}
  We have studied the electronic structure of the skutterudite
  compounds Co(Sb$_{1-x}$Te$_{x}$)$_3$ ($x=$ 0, 0.02, 0.04) by
  photoemission spectroscopy.  Valence-band spectra revealed that Sb
  5{\it p} states are dominant near the Fermi level and are hybridized
  with Co 3$d$ states just below it. The spectra of $p$-type \ptype
  are well reproduced by the band-structure calculation, which
  suggests that the effect of electron correlations is not strong in
  CoSb$_3$. When Te is substituted for Sb and $n$-type carriers are
  doped into CoSb$_3$, the spectra are shifted to higher binding
  energies as predicted by the rigid-band model. From this shift and
  the free-electron model for the conduction and valence bands, we
  have estimated the band gap of \ptype to be 0.03-0.04 eV, which is
  consistent with the result of transport measurements. Photoemission
  spectra of \Rh have also been measured and revealed similarities to
  and differences from those of CoSb$_3$.

\end{abstract}
}

\newcommand{\HeI}{He \footnotesize I \normalsize }
\newcommand{\HeII}{He \footnotesize II \normalsize }
\newcommand{\Co}{Co(Sb$_{1-x}$Te$_{x}$)$_3$ ($x=$ 0, 0.02, 0.04) }
\newcommand{\ptype}{CoSb$_3$ }
\newcommand{\Rh}{RhSb$_3$ }
\newcommand{\med}{Co(Sb$_{0.98}$Te$_{0.02}$)$_3$ }
\newcommand{\ntype}{Co(Sb$_{0.96}$Te$_{0.04}$)$_3$ }
\newcommand{\xCo}{CoSb$_3$ ($x=$ 0) }
\newcommand{\ef}{$E_F$ }

\narrowtext
\section{Introduction}
The family of compounds called skutterudites, with 
the general chemical formula 
{\it T}{\it X}$_{3}$ (${\it T}=$ Co, Rh, Ir; ${\it X}=$ 
pnictogen P, As, Sb), have recently received much attention
due to their potential for thermoelectric
applications.~\cite{thermo1,thermo2} The ability of thermoelectric
materials is defined as the thermoelectric 
figure of merit,
\begin{equation}
 Z = \frac{S^2 \sigma}{\kappa_e + \kappa_l} \propto \frac{m^{*3/2} \mu}{\kappa_l},
\end{equation}
where {\it S} is the Seebeck coefficient, $\sigma$ is the electrical
conductivity, $\kappa_e$ is the electronic thermal conductivity,
$\kappa_l$ is the lattice thermal conductivity, $m^{*}$ is the
effective mass, and $\mu$ is the carrier mobility.  Among the
skutterudite compounds, \ptype has attracted particular attention
because it has a high carrier mobility (for {\it p}-type) and a heavy
effective mass (for {\it n}-type).~\cite{mandrus,caillat,morelli}
Band-structure calculations have been performed by several groups.
Singh and Pickett~\cite{SP} suggested that its linear band dispersion
near the Fermi level ($E_F$) causes the unusual transport properties.
On the other hand, Sofo and Mahan~\cite{sofo} reported that its band
structure is typical of a narrow gap semiconductor. Moreover, Jung
{\it et al.}~\cite{jung} proposed the chemical picture that the top of
the valence band at the $\Gamma$ point in the Brillouin zone
corresponds to the antibonding combinations of the $\pi_4$ orbitals of
the Sb ring along a crystallographic axis. It is also reported that
the effective mass of the conduction band in the band-structure
calculations~\cite{sofo} is nearly by a factor of ten smaller than
that estimated from the transport experiments,~\cite{caillat} which
may indicate the existence of some effects not described by the
conventional band picture. X-ray photoemission (XPS) experiments on
CoAs$_3$, \ptype and RhSb$_3$ were performed~\cite{anno}, but the
energy resolution was not sufficient to investigate the electronic
structures near $E_F$.  Therefore, it is important to perform a
photoemission study of \ptype with higher energy resolution.

In this paper, we report on the results of a photoemission study of
Co(Sb$_{1-x}$Te$_{x}$)$_3$ ($x=$ 0, 0.02, 0.04) including ultraviolet
photoemission spectroscopy(UPS).  First, we show the valence-band
spectra of CoSb$_3$ ($x = 0$), which were taken with much higher
resolution than in the previous reports,~\cite{anno} and revealed that
Sb 5{\it p} orbitals are indeed dominant near \ef in CoSb$_3$.
Second, we compare the valence-band spectra of \ptype with a
theoretical spectrum derived from the band-structure
calculation.~\cite{sofo} The photoemission spectra are well reproduced
by the theoretical spectrum, which suggests that the effect of
electron correlations is not strong in CoSb$_3$. Third, we show a
rigid-band like spectral shift in \Co as a function of $x$. From the
spectral shift we have estimated a band gap of \ptype to be 0.03-0.04
eV. We also compare the spectra of \ptype with those of \Rh and
discuss the differences between them. Finally, the temperature
dependence of the spectra near \ef have been studied and have revealed
unusual behaviors in both compounds.

\section{Experiment}
Samples of \Co were prepared by the following method. Co (99.998\%
pure), Sb (99.9999\% pure) and Te (99.9999\% pure) powders were used
as starting materials. They were mixed in stoichiometric ratio \Co in
a plastic vial before being loaded in a steel die, where they were
compressed into a dense cylindrical pellet. The pellet was loaded on a
graphite boat in an Ar atmosphere, which was heated for two days at
823 K. The product was then crushed and ground in an alumina mortar.
Following the powder reaction, several samples of
Co(Sb$_{1-x}$Te$_{x}$)$_3$ were sintered by the spark plasma sintering
method. The spark plasma sintering was performed under a pressure of
30 MPa at 853 K for 10 min. \ptype was single-crystalline and showed
{\it p}-type transport, while \med and \ntype were polycrystalline and
showed {\it n}-type transport.

We performed the photoemission experiments using a hemispherical
electron energy analyzer. The light sources were the \HeI (${\it
  h}{\nu} = 21.2~$eV) and \HeII (${\it h}{\nu} = 40.8~$eV) resonance
lines for UPS, and the Mg K${\alpha}$ (${\it h}{\nu} = 1253.6~$eV)
line for XPS.  Energy calibration and estimation of the instrumental
resolution were done for Au film evaporated on the samples after each
series of measurements. The energy resolution was about 25 meV, 100
meV and 0.7 eV for \HeI UPS, \HeII UPS and XPS, respectively.
Measurements were done at 30 K, 100 K and 300 K for \HeI, and only at
300 K for \HeII and XPS. The base pressure in the spectrometer was
$\sim$ 1 $\times$ 10$^{-10}$ Torr for UPS and $\sim$ 5 $\times$
10$^{-10}$ Torr for XPS.  The surfaces of the samples were repeatedly
scraped {\it in situ} with a diamond file. When they were scraped at
low temperatures ($T <$ 100 K), however, the structures caused by the
adsorption of impurity were observed in the photoemission spectra.
Therefore, we measured only 30 K spectra after scraping at 100 K and
lowering the temperature down to 30 K.

\section{Results and Discussion}

\begin{figure}[ttt]
\begin{center}
\epsfile{file=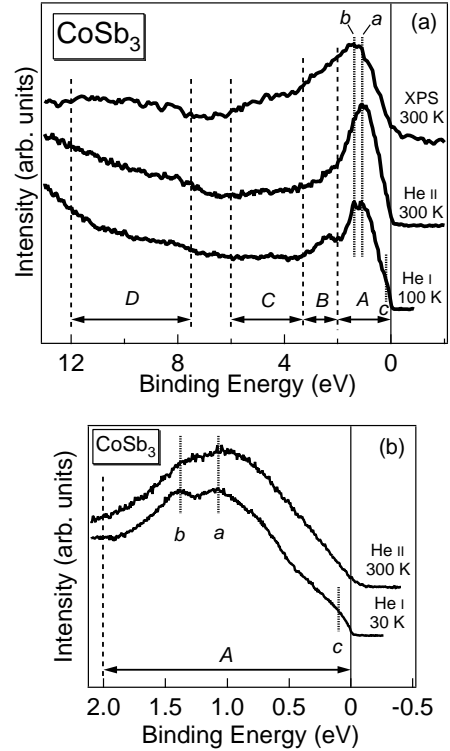,width=70mm}
\caption{Valence-band photoemission spectra of \ptype in the entire
  valence-band region (a) and in the vicinity of \ef (b).}
\label{cosbwn}
\end{center}
\end{figure}

\begin{figure}[ttt]
\begin{center}
\epsfile{file=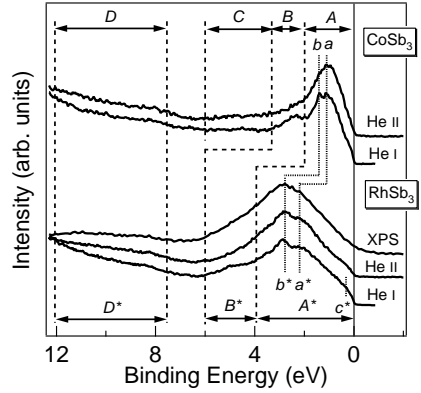,width=60mm}
\caption{Photoemission spectra of \Rh compared with those of
  CoSb$_3$.}
\label{rhsb}
\end{center}
\end{figure}

Figure~\ref{cosbwn}(a) and (b) show the valence-band photoemission
spectra of \ptype in the entire valence-band region and in the
vicinity of $E_F$, respectively, at various photon energies. Roughly
speaking, four structures are distinguishable in the spectra. Here,
these structures are referred to as {\it A} (from 0 to 2 eV), {\it B}
(from 2 to 3.3 eV), {\it C} (from 3.3 to 6 eV) and {\it D} (from 7.5
to 12 eV) as indicated in the figure. Based on the fact that the
relative cross-section of Sb 5{\it p} to Co 3{\it d} is the smallest
at ${\it h}{\nu} = 40.8~$eV,~\cite{cross} one can make the following
assignment of the spectral features. The distinct peak of {\it A} is
assigned to be mainly of Co 3{\it d} character because it is strong in
all the spectra. Peak {\it A} consists of two structures, {\it a} and
{\it b}, as indicated in Fig.~\ref{cosbwn}. {\it a} is observed both
in the \HeI and \HeII spectra, while {\it b} is hardly observed in the
\HeII spectrum. Judging from the relative cross-section of Sb 5{\it p}
to Co 3{\it d},~\cite{cross} one can assign {\it a} to relatively pure
Co 3{\it d} character and {\it b} to Co 3{\it d} states hybridized
with Sb 5{\it p} states.  Moreover, a small hump denoted by {\it c} is
observed in the region between 0 and 0.3 eV in the \HeI spectrum and
not in the \HeII spectrum. In a similar way to {\it b}, we assign this
hump to Sb 5{\it p} character. These results indicate that the
valence-band maximum has predominantly Sb 5{\it p} character,
consistent with the results of the previous band-structure
calculations.~\cite{SP,sofo,jung} Structures {\it B} and {\it C} can
be assigned to Sb 5{\it p} character, because they are observed in the
\HeI and XPS spectra but not in the \HeII spectrum.  According to the
band-structure calculation,~\cite{sofo} Sb 5{\it p} has a broad
structure in the region between \ef and 6 eV, and therefore {\it C} is
part of the broad structure. On the other hand, {\it B} includes Co
3{\it d} character, too, because, when we substitute Rh for Co, {\it
  C} does not change its binding energy, but {\it B} changes its
binding energy together with {\it A}, as described below
(Fig.~\ref{rhsb}). One can assign structure {\it D} to Sb 5{\it s}
character, because it increases with photon energy, which is
consistent with the fact that the relative cross-section of Sb 5{\it
  s} to Co 3{\it d} increases with photon energy.~\cite{cross}

\begin{figure}[htb]
\begin{center}
\epsfile{file=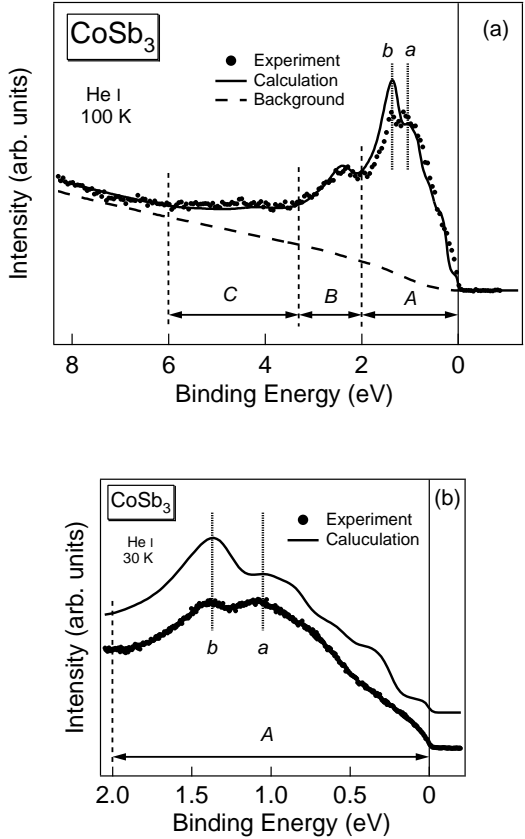,width=80mm}
\caption{Photoemission spectra of \ptype compared with the
  band-structure calculation~\cite{sofo} in the entire valence-band
  region (a) and in the vicinity of \ef (b). The dashed curve shows
  the background (see text). The spectra have been normalized to the
  area between \ef and 6 eV.}
\label{calcosb}
\end{center}
\end{figure}

In Fig.~\ref{calcosb}, we make a quantitative comparison between the
\HeI photoemission spectrum and the theoretical spectrum derived from
the band-structure calculation by Sofo and Mahan.~\cite{sofo} To
derive the theoretical photoemission spectrum we have taken into
account the contribution of the Co 3{\it d}, Sb 5{\it s} and Sb 5{\it
  p} partial density of states (DOS). They have been weighted by the
corresponding photoionization cross-sections at that photon
energy.~\cite{cross} This weighted DOS has been broadened by
convoluting with a Gaussian and a Lorentzian which represent the
instrumental resolution and the lifetime broadening, respectively.  We
assume that the lifetime width increases linearly with energy {\it E}
measured from $E_F$, i.e., FWHM $= \alpha |E - E_F|$.  The
coefficient $\alpha$, which represents the lifetime of the photo-hole
and increases with increasing binding energy, is empirically
determined to be $\sim$ 0.1.  Finally, we have added the background of
Henrich type~\cite{henrich} to the broadened DOS, as shown by a dashed
line in Fig.~\ref{calcosb}(a).  The spectra have been normalized to
the area between \ef and 7 eV.

In the wide energy range [Fig.~\ref{calcosb}(a)], the photoemission
spectrum is in good agreement with the theoretical spectrum.  The main
structures, {\it A}, {\it B}, and {\it C}, which were defined in
Fig.~\ref{cosbwn}, are well reproduced in the theoretical simulation.
These results suggest that electron correlations are not important in
CoSb$_3$.  However, there exists a small discrepancy in the fine
structures in {\it A} between the photoemission spectrum and the
theoretical spectrum.  The intensity ratio {\it b}/{\it a} is
different between experiment and calculation.  The intensity of {\it
  b} is as high as that of {\it a} in the photoemission spectrum,
while it is apparently higher than that of {\it a} in the theoretical
spectrum.  As mentioned above, {\it a} and {\it b} can be assigned to
relatively pure Co 3{\it d} and Co 3{\it d} states hybridizing with Sb
5{\it p} states, respectively.  The local-density approximation used
for the band-structure calculation tends to overestimate the
hybridization between the Co 3{\it d} and Sb 5{\it p} electrons, and
hence the hybridization strength may have been overestimated.

\begin{figure}[htb]
\begin{center}
\epsfile{file=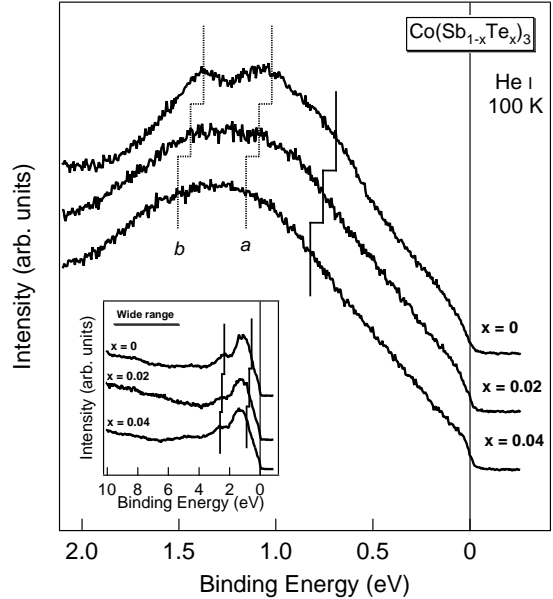,width=80mm}
\caption{Photoemission spectra of Co(Sb$_{1-x}$Te$_{x}$)$_3$ ($x=$ 0, 0.02,
  0.04) near $E_F$. The inset shows wide-range spectra. The spectra
  have been normalized to the height of the Co 3{\it d} peak.}
\label{cosbshift}
\end{center}
\end{figure}

\begin{figure}[htb]
\begin{center}
\epsfile{file=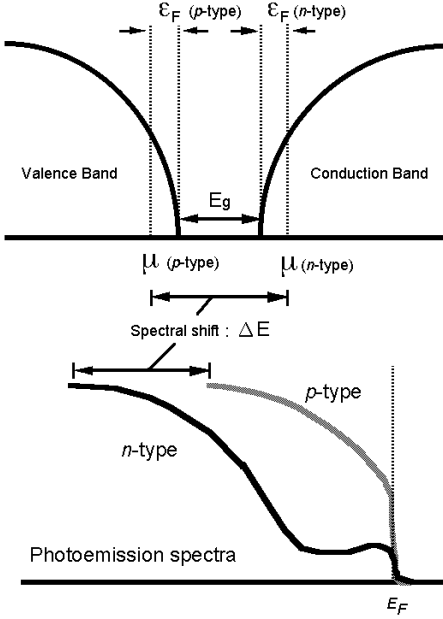,width=60mm}
\caption{Schematic illustration of the spectral DOS for {\it p}-type and
  {\it n}-type semiconductors, which is observed by photoemission
  spectroscopy.}
\label{shift}
\end{center}
\end{figure}

\begin{table}
 \caption[rigid]{Values for the Fermi energy $\epsilon_F$, the
   effective mass $m^*$, and the carrier concentration $n$. The values
   of $m^*$ are taken from Ref. ~\onlinecite{caillat} and $m_e$ is the
   free-electron mass.}
 \begin{center}
  \begin{tabular}{lrrrr} 
   \multicolumn{1}{c}{} &\multicolumn{1}{c}{type}&  \multicolumn{1}{c}{$\epsilon_F$} &  \multicolumn{1}{c}{$m^*$}&
   \multicolumn{1}{c}{$n$} \\ 
   \multicolumn{1}{c}{} &\multicolumn{1}{c}{}& \multicolumn{1}{c}{[eV]} &
   \multicolumn{1}{c}{} & \multicolumn{1}{c}{[cm$^{-3}$]} \\ \hline
    \ptype &{\it p}  & 0.03  & 0.15 $\times$ $m_e$ & 1.4 $\times$ 10$^{18}$  \\
    \ntype &{\it n}  & 0.06 & 3.5 $\times$ $m_e$ & 4 $\times$ 10$^{20}$    \\
  \end{tabular}
 \end{center}
 \label{rigid}
\end{table}%

Figure~\ref{cosbshift} shows the valence-band spectra of
Co(Sb$_{1-x}$Te$_x$)$_3$ in the vicinity of $E_F$.  The spectra have
been normalized to the height of the peak at $\sim$ 1.2 eV. The
spectra of the $x = 0.02$ and $x = 0.04$ samples are more blurred than
the $x = 0$ sample, probably because the substitution of Te for Sb
caused potential disorder in the crystal. Energy shifts of them were
determined by comparing the positions of the low binding energy side
of the Co 3{\it d} peak, as indicated by a solid line in the figure.
The spectra of Co(Sb$_{1-x}$Te$_x$)$_3$ are shifted to higher binding
energy with increasing $x$. This is what would be expected from the
rigid-band model since Te has one more valence electron than Sb and
hence each substituted Te atom donates oner electron to the conduction
band. We obtained the spectral shifts of 0.05-0.06, 0.06-0.07, and
0.12-0.13 eV for $x = 0 \to 0.02$, $x = 0.02 \to 0.04$, and $x = 0 \to
0.04$, respectively. From these shifts we estimated the band gap of
\ptype as follows. The $x = 0$ and $x = 0.04$ samples are {\it p}-type
and {\it n}-type degenerate semiconductors, respectively, and their
chemical potentials are located below the valence-band maximum for the
{\it p}-type material and above the conduction-band minimum for the
{\it n}-type material, as illustrated in Fig.~\ref{shift}.  In the
regions between $\mu$ and the band edge, one can regard the energy
band to be parabolic. Then the separation between $\mu$ and the band
edge corresponds to the Fermi energy $\epsilon_f$ of the carriers and
is given by:
\begin{equation}
 \epsilon_F = \frac{\hbar^2{k_F}^2}{2m^*},
\end{equation} 
where $k_F$ and $m^*$ are the Fermi wave vector and the effective mass of the
carriers, respectively. $k_F$ is given by:
\begin{equation}
 k_F = (3 \pi^2 n)^{1/3},
\end{equation} 
where $n$ is the carrier concentration. The values of $\epsilon_F$,
$k_F$, $m^*$, and $n$ in the measured samples are listed in
Table~\ref{rigid}. Here, we used the room temperature values of {\it
  n}, because the values at room temperature are little different from
those at 100 K in the case of degenerate semiconductors. In the
photoemission spectra, the binding energy corresponds to the energy
difference from $\mu$. When electrons are doped and $\mu$ rises, the
spectra are shifted to higher binding energies, as illustrated in
Fig.~\ref{shift}. In this way, one can estimate the band gap $E_g$
from the measured shift $\Delta E$ as:
\begin{equation}
 E_g = \Delta E - \epsilon_{F(x = 0)} - \epsilon_{F(x = 0.04)}
\end{equation} 
and obtains the value of 0.03-0.04 eV for $E_g$. This value is
consistent with the value of $E_g \sim 0.05$ eV deduced from the
transport experiments~\cite{mandrus,caillat,arushanov} within the
experimental uncertainties.

Figure~\ref{rhsb} shows the valence-band photoemission spectra of
RhSb$_3$.  In analogy to CoSb$_3$, we refer to the main structures as
{\it A$^*$} (from 0 to 4 eV), {\it B$^*$} (from 4 to 6 eV) and {\it
  D$^*$} (from 7 to 12 eV) as indicated in the figure, corresponding
to {\it A}, {\it B} and {\it D} in the \ptype spectra. Structure {\it
  C$^*$}, which should correspond to {\it C}, is not observed,
probably because it is hidden by {\it A$^*$} and {\it B$^*$}, which
are located at higher binding energies and are broader than those in
CoSb$_3$. Moreover, fine structures {\it a$^*$}, {\it b$^*$} and {\it
  c$^*$} within {\it A$^*$} are also defined as indicated in the
figure. As in the case of CoSb$_3$, structure {\it A$^*$} is assigned
to Rh 4{\it d}.  {\it A$^*$} is broader than {\it A}, which indicates
that the Rh 4{\it d} band is wider than the Co 3{\it d} band.
Moreover, the intensity ratio {\it b$^*$}/{\it a$^*$} is different
from {\it b}/{\it a} of CoSb$_3$. For CoSb$_3$, {\it b} is as strong
as {\it a} in the \HeI spectrum and is much weaker than {\it a} in the
\HeII spectrum. On the other hand, for RhSb$_3$, {\it b$^*$} is
stronger than {\it a$^*$} both in the \HeI and \HeII spectra, although
the intensity ratio {\it b$^*$}/{\it a$^*$} is slightly smaller in the
\HeII spectrum. One can explain these differences as follows. From the
results of the \ptype spectra, we can regard {\it a$^*$} as pure {\it
  d} character and {\it b$^*$} as hybridized {\it d} character. Since
Rh 4{\it d} electrons are more itinerant than Co 3{\it d} electrons,
{\it b$^*$}/{\it a$^*$} for \Rh is larger than CoSb$_3$. The hump just
below $E_F$, {\it c$^*$}, is also identified in RhSb$_3$ and can be
assigned to Sb 5{\it p} states. Structures {\it B$^*$} and {\it
  D$^*$} are assigned to Sb 5{\it p} character and to Sb 5{\it s}
character, respectively, in the same way as {\it B} and {\it D}.
Comparing the \Rh spectra with the \ptype spectra, {\it B$^*$} is
located at a higher binding energy than {\it B} in the same way as
{\it A$^*$} ({\it A}), which is of {\it d} character.  On the other
hand, structures {\it D$^*$} is located at the same binding energies
as {\it D}. {\it C$^*$} is probably hidden by {\it B$^*$} and, if this
is the case, it is located at the same energy as $C$.  These results
suggest that {\it B$^*$} ({\it B}) hybridizes with {\it d} states and
that {\it C$^*$} ({\it C}) does not.

\begin{figure}[ttt]
\begin{center}
\epsfile{file=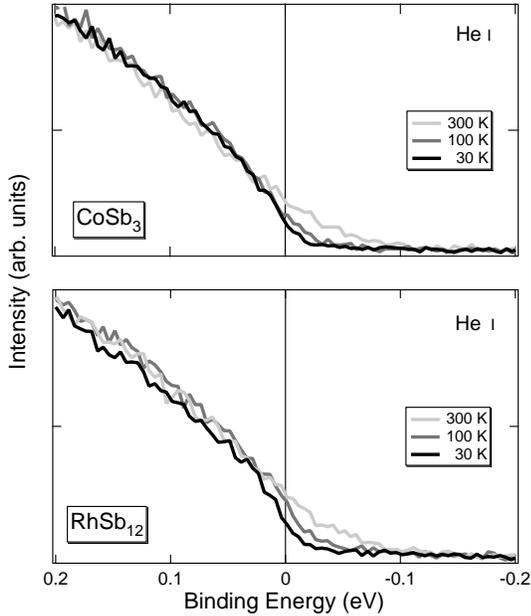,width=80mm}
\caption{Temperature-dependent photoemission spectra of \ptype and \Rh
  in the region
  near $E_F$. The spectra have been normalized to the area between
  -0.2 and 2 eV for CoSb$_3$, and between -0.2 and 4 eV for RhSb$_3$.}
\label{temp}
\end{center}
\end{figure}

\begin{figure}[tbh]
\begin{center}
\epsfile{file=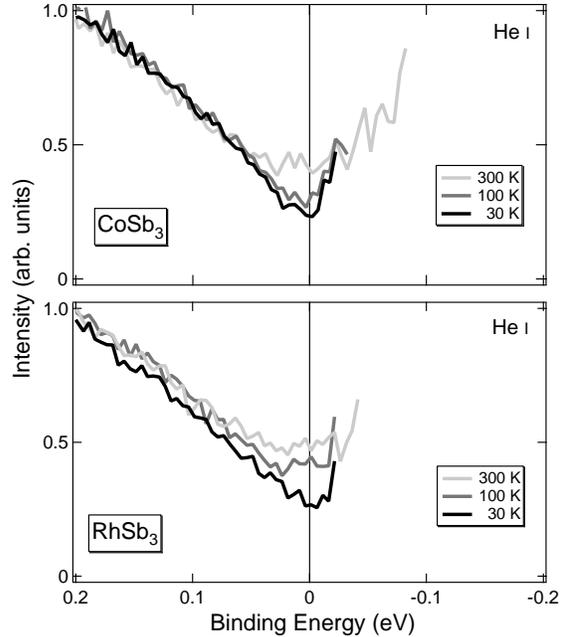,width=80mm}
\caption{Temperature-dependent spectral DOS, which have been obtained
  by dividing the spectra in Fig.~\ref{temp} by the Fermi Dirac
  distribution function.}
\label{divided}
\end{center}
\end{figure}

Finally, Fig.~\ref{temp} shows the temperature-dependent spectra of
\ptype and \Rh in the region near $E_F$. The spectra have been
normalized to the area between -0.1 and 2 eV for \ptype and between
-0.1 and 4 eV for RhSb$_3$. One can see that the intensities of the
spectra around \ef decrease with decreasing temperature.  To isolate
the temperature dependence of the spectral DOS from the temperature
dependence of the Fermi-Dirac (FD) distribution function, we have
divided the spectra by the FD distribution function (convoluted with a
Gaussian corresponding to the instrumental
resolution)~\cite{gap1,gap2} as indicated in Fig.~\ref{divided}.  In
the figure, we find in each spectrum the formation of a dip in the DOS
at $E_F$, which becomes deeper with decreasing temperature. The
magnitudes of these dips (pseudogaps) are $\sim$ 0.05 and $\sim$ 0.1
eV for \ptype and RhSb$_3$, respectively.  These results may indicate
spectral weight transfer over large energies of order eV. This
anomalous temperature dependence near $E_F$ cannot be explained by the
simple band picture.  Here we may suggest two candidates which can
cause the above phenomena. First, we may consider electron-electron
interaction between the Sb 5$p$ electrons forming the antibonding
combinations of the $\pi_4$ orbitals of the Sb ring, which was
reported by Jung {\it et al}.~\cite{jung} This possibility is,
however, rather unlikely because of the itinerant nature of the Sb
5$p$ electrons.  Second, electron-phonon interactions may be
considered. It has been reported that the most prominent lines in
Raman spectra of \ptype and RhSb$_3$ are due to the Sb$_4$-ring
breathing modes and their energy is $\sim$ 0.02 eV,~\cite{raman} which
may modulate the electronic states around the valence-band maximum
with decreasing temperature from 300 K. Indeed, Sofo and Mahan~\cite{sofo}
reported that small changes in the positions of the Sb atoms affect
strongly the electronic structures near $E_F$. Further studies are
necessary to elucidate the origin of the unusual observations.

\section{Conclusion}

We have performed the photoemission study of \Co and RhSb$_3$.  The
top of the valence band of \ptype is of Sb 5{\it p} character
hybridized with Co 3$d$ state just below it. The photoemission spectra
of \ptype were well reproduced by the band-structure
calculation~\cite{sofo}, which suggests that electron correlation does
not play an important role in the electronic structure of CoSb$_3$.
When Te is substituted for Sb and electrons are doped into CoSb$_3$,
the photoemission spectra are rigidly shifted to higher binding
energies.  From this spectral shift, we have estimated the band gap of
\ptype to be 0.03-0.04 eV, which is consistent with the transport
experiments.~\cite{mandrus,caillat,arushanov} The spectra of \Rh are
similar to CoSb$_3$ to some extent, whereas some differences are
caused by the wider band width of Rh 4{\it d}.  Unusual
temperature-dependence has been observed for the spectra of \ptype and
\Rh in the region near $E_F$, which we tentatively attribute to the
effect of Sb$_4$-ring breathing phonons.

\section{Acknowledgment}
The authors would like to thank T. Susaki, J. Matsuno and A. Chainani
for their useful discussions. One of us J. O. S. is supported by CONICET
Argentina.


\begin{references}
\bibitem{thermo1} B. C. Sales, D. Mandrus, and R. K. Williams, Science
  {\bf 272}, 1325 (1996).
\bibitem{thermo2} J.-P. Fleurial, T. Caillat, A. Borshchevsky,
  D. T. Morelli, and G. P. Meisner, in {\it Proceedings of the 15th
    International Conference on Thermoelectrics}, edited by T. Caillat
  (Institute of Electrical and Electronics Engineers, Piscataway, New Jersey,
  1996), p. 91.
\bibitem{mandrus} D. Mandrus, A. Migliori, T. W. Darling,
  M. F. Hundley, E. J. Peterson, and J. D. Thompson, Phys. Rev. B {\bf 
    52}, 4926 (1995).
\bibitem{caillat} T. Caillat, A. Borshchevsky, and J.-P. Fleurial,
  J. Appl. Phys. {\bf 80}, 4442 (1996).
\bibitem{morelli} D. T. Morelli, T. Caillat, J.-P. Fleurial,
  A. Borshchevsky, and J. Vandersande, B.Chen, and C. Uher, Phys. Rev.
  B {\bf 51} 9622 (1995).
\bibitem{SP} D. J. Singh and W. E. Pickett, Phys. Rev. B {\bf 50}, 11
  235 (1994).
\bibitem{sofo} J. O. Sofo and G. D. Mahan, Phys. Rev. B {\bf 58}, 15
  620 (1998).
\bibitem{jung} Dongwoon Jung, Myung-Hwan Whangbo, and Santiago
  Alvarez, Inorg. Chem. {\bf 29}, 2252 (1990).
\bibitem{anno} H. Anno, K. Matsubara, T. Caillat, and J.-P. Fleurial,
  Phys. Rev. B {\bf 62} 10 737 (2000).
\bibitem{cross} J.-J. Yeh and I. Lindau, At Data Nucl. Data Tables
  {\bf 32}, 1 (1985).
\bibitem{henrich} X. Li, Z. Zhang, and V. E. Henrich, J. Electron
  Spectrosc. Relat. Phenom. {\bf 63}, 253 (1993).
\bibitem{arushanov} E. Arushanov, M. Respaud, H. Rakoto, J. M. Broto,
  and T. Caillat, Phys. Rev. B {\bf 61}, 4672 (2000).
\bibitem{gap1} T. Susaki, Y. Takeda, M. Arita, K. Mamiya, A. Fujimori,
  K. Shimada, H. Namatame, M. Taniguchi, N. Shimizu, F. Iga, and
  T. Takabatake, Phys. Rev. Lett. {\bf 82}, 992 (1999).
\bibitem{gap2} H. Kumigashira, T. Sato, T. Yokota, T. Takahashi,
  S. Yoshii, and M. Kasuya, Phys. Rev. Lett. {\bf 82}, 1943 (1999).
\bibitem{raman} G. S. Nolas, G. A. Slack, T. Caillat, and
  G. P. Meisner, J. Appl. Phys. {\bf 79}, 2622 (1996).
\end{references}
\end{document}